\documentclass[10pt,journal]{IEEEtran}
\usepackage{soul}
\usepackage{dsfont}
\usepackage{amsmath}
\usepackage{msc}
\usepackage{tikz}
\usetikzlibrary{shapes,arrows}
\usepackage{multirow}
\usepackage{balance}
\usepackage{cite}
\usepackage{setspace,amsmath,latexsym,cite,amssymb,epsfig,amsfonts}
\usepackage{url,cite}
\usepackage{balance}
\usepackage{graphicx}
\usepackage{enumerate}
\usepackage{stmaryrd}
\usepackage{colortbl}
\usepackage{tcolorbox}

\ifCLASSINFOpdf
\else
\fi
\begin{document}
%
\title{Scalar Product Lattice Computation for Efficient  Privacy-preserving Systems}
\author{Yogachandran Rahulamathavan,  Safak Dogan, \textit{Senior Member, IEEE}, Xiyu Shi, \textit{Member, IEEE}, \\Rongxing Lu, \textit{Senior Member, IEEE}, Muttukrishnan Rajarajan, \textit{Senior Member, IEEE}, \\and Ahmet Kondoz, \textit{Senior Member, IEEE}
\IEEEcompsocitemizethanks{ 
\IEEEcompsocthanksitem Y. Rahulamathavan,  S. Dogan, X. Shi and A. Kondoz are with the Institute for Digital Technologies, Loughborough University London, London, U.K. (e-mails: \{y.rahulamathavan, s.dogan, x.shi, a.kondoz\}@lboro.ac.uk).
 \IEEEcompsocthanksitem R. Lu is with the Faculty of Computer Science, University of New
 Brunswick, Fredericton, NB E3B 5A3, Canada.  (e-mail:  RLU1@unb.ca).
  \IEEEcompsocthanksitem M. Rajarajan is with the Information Security Group, School of Engineering and Mathematical Sciences, City University London,
  EC1V 0HB, London, U.K. (e-mails: R.Muttukrishnan@city.ac.uk).
\IEEEcompsocthanksitem The work was supported by  UK-India Education Research Initiative (UKIERI) through grant UGC-UKIERI-2016-17-019.
\IEEEcompsocthanksitem {{Source code for this work can be found in Github repo (https://github.com/rahulay1/LWE)}}
}
}

%
%

\markboth{IEEE TRANSACTIONS ON }%
{Shell \MakeLowercase{\textit{et al.}}: Bare Demo of IEEEtran.cls for IEEE Journals}
%




\IEEEcompsoctitleabstractindextext{
\begin{abstract}
Privacy-preserving  applications allow users to perform on-line daily actions without leaking sensitive information. The privacy-preserving scalar product is one of the  critical algorithms in many private applications. The state-of-the-art privacy-preserving scalar product schemes use either computationally intensive homomorphic (public-key) encryption techniques such as Paillier encryption to achieve strong security (i.e., $128-$bit) or random masking technique to achieve high efficiency for low security. In this paper, lattice structures have been exploited to develop an efficient privacy-preserving system. The proposed scheme is not only efficient in computation as compared to the state-of-the-art  but also provides high degree of security against quantum attacks. Rigorous security and privacy analyses of the proposed scheme have been provided along with a concrete set of  parameters to achieve $128-$bit and $256-bit$ security. Performance analysis shows that the scheme is at least five orders faster than the Paillier schemes and at least twice as faster than the existing randomisation technique at $128-$bit security. {{Also the proposed scheme requires six-time fewer data compared to Paillier and randomisation based schemes for communications.}}
\end{abstract}

}

%
\maketitle
\IEEEdisplaynotcompsoctitleabstractindextext
\IEEEpeerreviewmaketitle

\section{Introduction}
\IEEEPARstart{R}{egulators}  around the world are enforcing privacy-by-design and privacy-by-default approaches to protect the users' data in rest, transit and processing. Several service providers and applications that traditionally use users' data in plain domain to extract patterns and provide services are now applying encrypted domain computations. Some of the example applications are disease classification in health-care, data search in the cloud, biometric verification, etc. (e.g., \cite{Barni2013, Barni2011, Barni2009,Rahul_TASL, Rahul_TDSC_2016,Rahul_TDSC,Rahul_Bio,Rahul_TAFF} and references their-in).
The common theme across these applications is that there are two distrusting parties want to work on a common goal by combining both of their data while preserving the data privacy. For example, a buyer wants to verify his age to an on-line shop using security token instead of sending date of birth.

There are algorithms developed in literature  to support data privacy for  applications such as classification algorithms, data mining algorithms, distance calculations etc. \cite{Barni2013, Barni2011, Barni2009, Rahul_TASL, Rahul_TDSC_2016,Rahul_TDSC,Rahul_Bio,Rahul_TAFF}. In all of these algorithms, one party  encrypts the sensitive data whenever that data should be sent to other party. Hence the second party needs to process the received data in an encrypted domain. This approach ensures data privacy.   Regardless of algorithms,  privacy-preserving scalar product (PPSP)  has been used as one of the privacy enabling tools between the two parties. The intuition behind this is that a mathematical function that relies on two different variables can be modified into a scalar product \cite{ Barni2009, Rahul_TASL}. Therefore, PPSP becomes a vital tool in most of the privacy-preserving (PP) algorithms. 

Suppose, there are two parties, A and B, want to compute the following scalar product
\begin{equation*}
\mathbf{a}^T\mathbf{b} = \sum_{i=1}^{n}a_i.b_i,
\end{equation*}
where vector $\mathbf{a} = (a_1, a_2, \ldots, a_n)$ belongs to  A and vector $\mathbf{b} = (b_1, b_2, \ldots, b_n)$ belongs to  B. The privacy requirement here is that no party is allowed to learn the others input vector. At the end, only one party can learn the output of the scalar product (SP). 

Several solutions have been proposed to address this problem in literature (see Section \ref{S: Literature Survey}). These solutions  rely on either public-key encryption techniques to achieve strong security or  randomisation techniques for high efficiency. The security of these schemes rely on mathematically hard problems and these solutions will be obsolete in few years time due to the rise of quantum computers as there are existing quantum algorithms which can easily  solve the  mathematically intractable problems \cite{PerkertLatticeforInternet,PerkertLatticeDecadeofLattice,RegevLWE2005,RegevLWE2009,Ajtai1996}. 

Hence, this paper exploits \textit{lattice-based cryptography} to build a PPSP. The proposed model is similar to lattice-based fully homomorphic encryption scheme \cite{RegevLWE2005} and support multiple encryption and addition without decryption \cite{PerkertLatticeforInternet}. However, the major challenge was to ensure the error terms are not overflowed to effect the accuracy. The paper proposes a methodology to control the error terms while ensuring the given security level, i.e., 128-bit.

Lattice-based cryptography has been proven to be secure against quantum attacks and expected to replace the existing public-key cryptography schemes \cite{PerkertLatticeforInternet,PerkertLatticeDecadeofLattice,RegevLWE2005,RegevLWE2009,Ajtai1996}. Therefore the proposed solution will be secure against quantum computers and can be used in PP algorithms for various applications to achieve privacy. At the same time, the experimental results (see Section \ref{Section: Experiments}) show that the proposed PPSP can also be executed significantly faster than the existing PPSP schemes at equivalent security level.

The rest of this paper is organised as follows:  The related work is discussed in Section \ref{S: Literature Survey}. The background information about lattice-based cryptography and its hardness assumptions are provided in Section \ref{Section: Intro Lattice}. The proposed algorithm is described in Section \ref{S: Binary PPSP}  followed by the security analysis and parameter selections in Section \ref{S: Security Analysis}. Experimental results are provided in Section \ref{Section: Experiments}. The conclusions and future work are discussed in Section \ref{Section: Conclusions}. 

\section{Literature review}\label{S: Literature Survey}
The existing PPSP schemes can broadly be divided into two: 1) the schemes that are built using proven cryptography such as homomorphic encryption, and 2) the schemes that are built based on information theory such as randomisation  and linear algebra. Even though the latter is much efficient than former, security level of latter is not quantified. The following subsections study the state-of-the-art algorithms for each of these schemes.
\subsection{Homomorphic encryption based PPSP} \label{SSSection: HE based Scalar product}
Homomorphic encryption techniques such as Paillier play a vital role in supporting PPSP since it offers high security such as $128-$bits \cite{Paillier}. Even though this scheme is highly secure, it  becomes inefficient with the size of the vectors i.e., it may take long time (i.e., a few minutes in modern laptops with five cores and 6GB memory) to compute the scalar product when the dimension of the vectors is around 1000. Several efficient PPSP schemes were proposed in literature to improve the efficiency \cite{ScalarPDu2001,ScalarPDu2002,ScalarPVaidya2002,ScalarPAmirbekyan2007,ScalarP1,ScalarP2,ScalarP3,SPOC2,LinearAlgebraScalarProduct}. All these schemes use the homomorphic PPSP scheme as a benchmark to measure the efficiency. We discuss these in the following subsections.     

A lattice based functional encryption technique that \textit{predicates} whether the SP is equivalent to $0$ or not $0$ was proposed in \cite{AgrawalLatticeScalar2011}. This work is based on lattice trapdoors \cite{MicciancioTrapdoor}. If the SP is equivalent to $0$ then the trapdoors successfully remove large elements in the problem. Note that the work in \cite{AgrawalLatticeScalar2011} is completely different to the objective of the proposed work on this paper and the algorithm in  \cite{AgrawalLatticeScalar2011} cannot be modified to develop a PPSP scheme.

There are works that directly uses Learning with errors based cryptographic scheme for encrypted domain matrix calculations \cite{HElib, MLConfidential, SmartGrid, DoingtheRealWork}. These works treat the encryption technique as a black-box to develop several applications ranging from logistic regression based prediction to statistics of smart meter reading in encrypted domain. In contrast to traditional homomorphic encryption such as Paillier, the  learning with error based encryption involve a number of parameters that must be set properly for problems with different dimensions. Otherwise, as we will show in Section \ref{Section: Intro Lattice}, error terms will overflow and decryption will be unsuccessful. In this paper, we  clearly show how to setup the parameters to achieve different level of security. \textbf{Most importantly this is the first paper that compares the performance of quantum secure cryptographic scheme against traditional homomorphic encryption scheme and information theoretic secure scheme and show that a quantum cryptographic based scheme can outperform the other schemes if the parameters are set properly.}

\subsection{Information theory based PPSP} \label{SSSection: Randomisation based Scalar product}
In 2001, Du et al proposed a PPSP algorithm using 1-out-of-N oblivious transfer function and homomorphic encryption \cite{ScalarPDu2001}. This algorithm is based on splitting the  input vector $\mathbf{a}$ of Party A into $p$ number of random vectors to achieve privacy from Party B. The drawback of this method is that both parties need to be on-line and interact several times to perform the SP. 

In 2002, Du et al proposed another SP which reduces the communication complexity of their previous work \cite{ScalarPDu2001} but with the help of a third-party semi-trusted server \cite{ScalarPDu2002}. The algorithm in  \cite{ScalarPDu2002} requires a third-party sever to generate two random vectors $\mathbf{R}_A$ and $\mathbf{R}_B$. The vector   $\mathbf{R}_A$ will be revealed to  A and the vector $\mathbf{R}_B$ will be revealed to  B. Using these vectors,  A and  B compute the shares of the SP. Hence, both the parties must reveal their shares to get the actual SP value. The communication complexity of this protocol is four times higher than the communication cost of SP  without privacy. Moreover, the major draw back of this work is the involvement of third-party  who can easily collude with one of the parties to reveal the other party's input vector.

Vaidya and Clifton in 2002 proposed a novel PPSP solution but without the need of third-party in \cite{ScalarPVaidya2002}. The communication complexity of the algorithm in \cite{ScalarPVaidya2002} is same as \cite{ScalarPDu2002}. However, the computation cost is $O(n^2)$ while it is $O(n)$ for the \cite{ScalarPDu2002}. Moreover, the security of the SP algorithm in \cite{ScalarPVaidya2002} depends on the difficulty of solving $n/2$ linear equations.

In 2007, Amirbekyan et. al. proposed a homomorphic encryption and randomisation (or add vector protocol) based PPSP \cite{ScalarPAmirbekyan2007}.   Since $2\mathbf{a}^T.\mathbf{b} = \sum_{i=1}^{n}a_i^2 + \sum_{i=1}^{n}b_i^2 -(\mathbf{a} -\mathbf{b})^2$, the authors of \cite{ScalarPAmirbekyan2007} exploited homomorphic encryption technique to compute $\mathbf{a} -\mathbf{b}$. Party A generates public and private key pairs using any homomorphic encryption scheme that offers additive homomorphism (i.e., Pailler encryption) and encrypt the elements of vector $\mathbf{a}$.  The encrypted vector and the public key are sent to Party B. Party B subtract its vector $\mathbf{b}$ from encrypted $\mathbf{a}$ using homomorphic properties and obtain encrypted $(\mathbf{a}-\mathbf{b})$. Subsequently, Party B permutes and sends the elements of encrypted $(\mathbf{a}-\mathbf{b})$ to Party A. Party A decrypts the vector received from Party B and obtains the permuted $(\mathbf{a}-\mathbf{b})$. Party A also receives $\sum_{i=1}^{n}b_i^2$ from Party B. Using these, Party A can compute the required SP. Similarly, there are several variations of PPSP algorithms  proposed in literature they either use homomorphic encryption  or randomisation or both  \cite{ScalarP1,ScalarP2,ScalarP3}.  

One of the algorithms that is secure and lightweight to-date is called Secure and Privacy-preserving Opportunistic Computing proposed in \cite{SPOC2} which is proven to be faster than  all the other SP and achieve high security. In \cite{SPOC2}, the security and privacy of the input vectors are protected by masking them by large random integers whose size is around $512$ bits.  It is shown in \cite{SPOC2}, that the computational complexity is almost negligible and communication complexity is almost half compared to the Paillier homomorphic encryption based SP \cite{Paillier}. To make a fair comparison with the proposed scheme, we reset the parameters to achieve $128-$bit security against traditional computers. Then in Section \ref{Section: Experiments},  we compare the performance of \cite{SPOC2} against the proposed lattice-based PPSP scheme and show that the latter one is, at least twice as fast as the \cite{SPOC2} algorithm.

Recently, linear algebra based PPSP  was proposed in \cite{LinearAlgebraScalarProduct} for biometric identification. The solution proposed  is efficient  and do not require  parties to be on-line. In particular, the solution is very useful when Party A wants to outsource the SP computation to Party B. 

For this scheme, Party A holds both the input vectors $\mathbf{a}$ and $\mathbf{b}$. Initially,  Party A obtains a diagonal matrix $\mathbf{A}$ using the input vector $\mathbf{a}$ followed by  generating two random invertable matrices $\mathbf{M}_1$ and $\mathbf{M}_2$ and a random lower triangular matrix $\mathbf{U}$. The encryption of the input vector $\mathbf{a}$ is simply a matrix multiplication i.e., $\mathbf{M}_1\mathbf{U}\mathbf{A}\mathbf{M}_2$. This encrypted matrix is send to  Party B. Later, if Party A wants to compute a SP $\mathbf{a}^T\mathbf{b}$ then Party A generates a random lower triangular matrix $\mathbf{V}$ and computes $\mathbf{M}_1^{-1}\mathbf{V}\mathbf{B}\mathbf{M}_1^{-1}$ as an encryption of $\mathbf{b}$ where matrix $\mathbf{B}$ is just a diagonal matrix of $\mathbf{b}$. This encrypted matrix is sent to Party B who computes the following which is equivalent to $\mathbf{a}^T\mathbf{b}$: $\mathbf{Tr}\{\mathbf{M}_1^{-1}\mathbf{V}\mathbf{B}\mathbf{M}_2^{-1} . \mathbf{M}_1\mathbf{U}\mathbf{A}\mathbf{M}_2\}$ where $\mathbf{Tr}$ is a matrix trace operation \cite{MatrixBook}.

This model has been applied in various biometric authentication applications. For example, recently, the work in \cite{OutsourcedBiometric} exploited this scheme to protect biometric templates. In \cite{OutsourcedBiometric}, the user extracts biometric template $\mathbf{a}$ and encrypts using random matrices as explained in the previous paragraph. Later, if the user wants to authenticate to the server, then the user extracts a new biometric sample, lets say $\mathbf{b}$, and encrypts using the random matrices and send it to server. Using these encrypted samples (i.e., $\mathbf{a}$  and $\mathbf{b}$), the server can find the similarities.   This model requires multiplication of several matrices and the complexity will increase substantially when the elements of the matrices are set to large integers to achieve $128-$bit or higher security.  Again, the security of these schemes are dependent on integer factorisation and vulnerable for quantum algorithms.


\section{Lattice based Cryptography}\label{Section: Intro Lattice}
\subsubsection*{Notations} {{ We use bold lower-case letters like  $\mathbf{x}$ to denote column vectors; for row vectors we use the transpose $\mathbf{x}^T$.}} We use bold upper-case letters like $\mathbf{A}$ to denote matrices, and identify a matrix with its ordered set of column vectors. We denote horizontal concatenation of vectors and/or matrices using vertical bar, e.g., $[\mathbf{A} | \mathbf{A.x}]$ {{where $.$ denotes the matrix multiplication}}. For any integer $q \geq 2$, we use $\mathds{Z}_q$ to denote the ring of integers  modulo $q$, $\mathds{Z}_q^{n\times m}$ to denote the set of $n\times m$ matrix with entries in $\mathds{Z}_q$. We denote a real number $x$ as $x \in \mathds{R}$.

\subsection{Lattices}
An \textit{$m-$dimensional lattice} $\Lambda$ is a full-rank discrete subgroup of $\mathds{R}^m$ \cite{PerkertLatticeDecadeofLattice}. Let $\mathbf{b}_1, ~\mathbf{b}_2,\ldots, \mathbf{b}_n$ denote the $n$ linearly independent vectors in $\mathds{R}^m$. Then  \textit{$m-$dimensional lattice} $\Lambda$ is defined to be the set of all integer combinations of $\mathbf{b}_1, ~\mathbf{b}_2,\ldots, \mathbf{b}_n$ as follows:
\begin{equation}
\Lambda = \sum_{i=1}^{n}x_i\mathbf{b}_i,
\end{equation}
where $x_i \in \mathds{Z}, \forall i$. The set of vectors $\mathbf{b}_1, ~\mathbf{b}_2,\ldots, \mathbf{b}_n$ is called \textit{basis} for the lattice $\Lambda$, and $n$ is called the rank of the lattice.
 
Without loss of generality, we consider \textit{integer lattices} i.e., whose points have coordinates in $\mathds{Z}^m$. Among these lattices, many cryptographic applications use a particular family of so-called ``$q-$ary'' integer lattices which contain $q\mathds{Z}^m$ as a sub-lattice for some small integer $q$. There are two different $q-$ary lattices  considered in many lattice-based cryptographic applications. Let us define them as follows:
\subsubsection{$\Lambda_q^{\bot}(A) $}
For instance,  for any integer $q \geq 2$ and any $\mathbf{A} \in \mathds{Z}_q^{n\times m}$, a set of vectors $\mathbf{e}\in \mathds{Z}^m$  that satisfy the following equation 
\begin{equation}
\mathbf{A}.\mathbf{e} = \mathbf{0}~ mod~ q
\end{equation}
forms a lattice of dimension $m$, which is closed under congruence modulo $q$. This lattice is denoted by $\Lambda_q^{\bot}(A)$ where 
\begin{equation}
\Lambda_q^{\bot}(A) := \{\mathbf{e} \in \mathds{Z}^m | \mathbf{A}.\mathbf{e} = \mathbf{0}~ mod~ q\}.
\end{equation}
Using $\Lambda_q^{\bot}(A)$, we define a  coset or shifted lattice $\Lambda_q^{\mathbf{u}}(A)$ where
\begin{eqnarray}
\nonumber \Lambda_q^{\mathbf{u}}(A) &:=& \{\mathbf{e} \in \mathds{Z}^m | \mathbf{A}.\mathbf{e} = \mathbf{u}~ mod~ q\},\\
&=& \Lambda_q^{\bot}(A) + \mathbf{x},
\end{eqnarray}
where $\mathbf{u} \in \mathds{Z}^n_q$ is an integer solution to
\begin{equation}
\mathbf{A}.\mathbf{x} =\mathbf{u}  ~ mod~ q.
\end{equation}

\subsubsection{$\Lambda(A^T)$} 
Similarly, we can define another $m-$dimensional \textit{q-ary} lattice, $\Lambda(A^T)$. For a set of vectors $\mathbf{e}\in \mathds{Z}^m$, and  $\mathbf{s}\in \mathds{Z}^n_q$ which satisfy the following equation:
\begin{equation}
\mathbf{e} = \mathbf{A}^T.\mathbf{s} ~ mod~ q
\end{equation}
where 
\begin{equation}
\Lambda(A^T) := \{\mathbf{e} \in \mathds{Z}^m | \mathbf{s} \in \mathds{Z}_q^n ~s.t.~ \mathbf{e} = \mathbf{A}^T.\mathbf{s} ~ mod~ q\}.
\end{equation}
It is easy to check that $\Lambda_q^{\bot}(A)$  and $\Lambda(A^T)$ are dual lattices. 

\subsection{Lattice Hard Problems}\label{Section: Lattice Hard Problems}
There are three well-known hard problems in lattice that  have been exploited by researchers to build several cryptographic applications. This section defines these hard problems briefly.

\subsubsection{Short integer solution} 
Hardness of finding a short integer solution (SIS)  was first exploited by Ajtai \cite{Ajtai1996}. The SIS has served as a foundation for many cryptographic applications such as one-way hash function, identification scheme and digital signature using lattices. The SIS can be defined as follows:
\begin{tcolorbox}
\subsubsection*{Definition for SIS} For a given $m$ uniformly random vectors $\mathbf{a}_i \in \mathds{Z}_q^{n}$, forming columns of a matrix $\mathbf{A} \in \mathds{Z}_q^{n\times m}$, finding a non-zero \textit{short} integer vector $\mathbf{z} \in \mathds{Z}^{m}$ with norm $||\mathbf{z}|| < \beta$ such that 
\begin{equation*}
\mathbf{Az} = \sum_{i=1}^{m}\mathbf{a}_i.z_i = \mathbf{0} ~mod~q~
\end{equation*}
is intractable.
\end{tcolorbox}
This problem has the following useful observations:
\begin{enumerate}
\item Without the requirement of $||\mathbf{z}|| < \beta$ i.e., ``short'' solution, it is easy to find a vector $\mathbf{z}$ via Gaussian elimination that satisfies $\mathbf{Az} = \mathbf{0} ~mod~q$.
\item The problem becomes easier to solve if  $m$ is increased  and difficult to solve if  $n$ is increased.
\item The norm bound $\beta$ and the number $m$ of the column vectors must be large enough that a solution is guaranteed to exist. This is the case when $\beta > \sqrt{n.log(q)}$.
\end{enumerate}

\subsubsection{Inhomogeneous short integer solution}
Inhomogeneous short integer solution (ISIS) is a variant of SIS. ISIS problem can be defined as follows \cite{PerkertLatticeforInternet,PerkertLatticeDecadeofLattice}:
\begin{tcolorbox}
\subsubsection*{Definition for ISIS} For a given $m$ uniformly random vectors $\mathbf{a}_i \in \mathds{Z}_q^{n}$, forming  columns of a matrix $\mathbf{A} \in \mathds{Z}_q^{n\times m}$, and a uniform random vector $\mathbf{u} \in \mathds{Z}_q^{n}$, finding a non-zero integer vector $\mathbf{z} \in \mathds{Z}^{m}$ with norm $||\mathbf{z}|| < \beta$ such that 
\begin{equation*}
\mathbf{Az} = \sum_{i=1}^{m}\mathbf{a}_i.z_i = \mathbf{u} ~mod~q~
\end{equation*}
is intractable. 
\end{tcolorbox}
\subsubsection{Learning with errors}\label{Subsection: LWE in intro}
Learning with errors (LWE)  \cite{RegevLWE2009,RegevLWE2005} is an encryption-enabling lattice-based problem but similar to SIS. To enable encryption, the LWE problem depends on a ``small" error distribution over integers. The LWE is parametrised by positive integers $n$  and $q$, and a small error distribution $\mathcal{X} \in \mathds{Z}_q$, which is typically be a ``rounded'' normal distribution with mean $0$ and standard deviation $\frac{\alpha q}{2\pi}$. The constant $\alpha$ plays a critical role in the security of LWE and it should be chosen as large as possible while satisfying the following condition \cite{RegevLWE2005}:
\begin{equation}\label{Equation: condition 1 for alpha}
\alpha q > 2\sqrt{n}.
\end{equation}
There are two versions of LWE based problems. Before defining these, let us define a distribution called \textit{LWE-distribution} as follows:

\begin{tcolorbox}
\subsubsection*{LWE Distribution}  For a given \textit{secret} vector $\mathbf{s} \in \mathds{Z}_q^n$, a sample from LWE distribution $\mathcal{A}_{\mathbf{s},\mathcal{X}} \in \mathds{Z}_q^n \times \mathds{Z}_q$ is obtained by choosing a vector $\mathbf{a} \in \mathds{Z}_q^n$ uniformly at random, a ``small" error $e \in \mathcal{X}$, and outputting $(\mathbf{a}, b = \mathbf{s}^T\mathbf{a} + e ~mod~q)$.
\end{tcolorbox}

\noindent Using the  LWE distribution, we can define two versions of LWE problem as follows:
\begin{tcolorbox}

\subsubsection*{1. Search-LWE} Given $m$ independent samples $(\mathbf{a}_i,b_i) \in \mathds{Z}_q^n \times \mathds{Z}_q$ drawn from the above LWE distribution $\mathcal{A}_{\mathbf{s},\mathcal{X}}$ for a uniformly random $\mathbf{s} \in \mathds{Z}_q^n $ (fixed for all samples), it is intractable to find $\mathbf{s}$.\\

\subsubsection*{2. Decision-LWE} Given $m$ independent samples $(\mathbf{a}_i,b_i) \in \mathds{Z}_q^n \times \mathds{Z}_q$ where every sample is distributed according to either: (1) $\mathcal{A}_{\mathbf{s},\mathcal{X}}$ for a uniformly random $\mathbf{s} \in \mathds{Z}_q^n $ (fixed for all samples), or (2) the uniform distribution, then distinguishing which is the case is intractable.
\end{tcolorbox}
We can have  the following observations from the two LWE problems outlined above:
\begin{enumerate}
\item Without the error term $e \in \mathcal{X}$,  the search-LWE problem can be solved easily using Gaussian elimination technique and  the secret $\mathbf{s}$ can be recovered.
\item Similarly for decision-LWE problem, without the error term $e \in \mathcal{X}$, Gaussian elimination technique will reveal with high probability that no solution $\mathbf{s}$  exists if it is not sampled from LWE distribution. 
\item If there are $m$ LWE samples $(\mathbf{a}_i,b_i) \leftarrow \mathcal{A}_{\mathbf{s},\mathcal{X}}$ for a uniformly random $\mathbf{s} \in \mathds{Z}_q^n $ (fixed for all samples), we can combine all $\mathbf{a}_i$s into a matrix $\mathbf{A} = [\mathbf{a}_1,\mathbf{a}_2,\ldots,\mathbf{a}_m] \in \mathds{Z}_q^{n \times m}$, $b_i$s into a vector $\mathbf{b}=[b_1, b_2, \ldots, b_m]^T$, and $e_i$s into a vector $\mathbf{e}=[e_1, e_2, \ldots, e_m]^T$ into the following vector-matrix linear equation
\begin{equation*}
\mathbf{b}^T=\mathbf{s}^T\mathbf{A}+\mathbf{e}^T ~(mod~q).
\end{equation*}
\end{enumerate}
In the following sections, we will exploit the above lattice hard problems  to develop the the lattice-based PPSP. 

\tikzstyle{Bigblock} = [rectangle, draw, fill=blue!20, 
    text width=10em, text centered, rounded corners, minimum height=4em]
\tikzstyle{block} = [rectangle, draw, fill=blue!20, align=left,
    text width=8em, text centered, rounded corners, minimum height=2em]
\tikzstyle{blocklong} = [rectangle, draw, fill=blue!20, align=left,
    text width=8em, text centered, rounded corners,node distance=3cm, minimum height=2em]
\tikzstyle{line} = [draw, -latex']
\tikzstyle{linenoarrow} = [draw, loosely dashed]
\tikzstyle{cloud} = [draw, circle,fill=red!20, node distance=3cm,
    minimum height=2em]
\tikzstyle{dot} = [draw, circle ,fill=red!20, node distance=1.3cm,
    minimum height=1pt]
    \tikzstyle{dotlong} = [draw, circle ,fill=red!20, node distance=170pt,
    minimum height=1pt]

\section{Lattice-based PP Scalar Product Computation}\label{S: Binary PPSP}
\begin{figure}
\centering
\begin{tikzpicture}[node distance = 2cm, auto]
    \node [Bigblock] (init) {\underline{System Initilisation}\\ Public Parameters\\
    $\mathbf{A}  \in \mathds{Z}_q^{n \times m}$};
    \node [cloud, left of=init] (EntityX) {Entity X};
    \node [cloud, right of=init] (EntityY) {Entity Y};
    \node [block, below of=EntityX] (dot1) {\underline{Inputs:}\\ $\mathbf{x} \in \{0,1\}^m$};
    \node [block, below of=EntityY] (dot2) {\underline{Inputs:}\\ $\mathbf{y} \in \{0,1\}^m$\\ $\mathbf{t} \in \mathds{Z}_q^{n}$,\\ $e_1  \leftarrow \mathcal{X}$,  $\mathbf{e}_2  \leftarrow \mathcal{X}^m$.};
    \node [block, below of=dot1] (Step1X) {\underline{Step 1:} \\$\mathbf{u} = \mathbf{Ax} \in \mathds{Z}_q^{n}$};
   \node [blocklong, below of=dot2] (Step2Y) {\underline{Step 2:} \\$c_1 = \mathbf{t}^T\mathbf{u}+ e_1  \in \mathds{Z}_q,$\\
   $\mathbf{c}_2^T = \mathbf{t}^T\mathbf{A}+ \mathbf{e}_2^T + \lfloor \frac{q}{m}\rceil\mathbf{y}^T  \in \mathds{Z}_q^{1 \times m}$};
    \node [block, below of=Step1X] (Step3X) {\underline{Step 3:} \\$s=\left\lfloor\frac{\mathbf{c}_2^T\mathbf{x} - c_1}{\lfloor \frac{q}{m}\rceil}\right\rceil$};
    \node [dot, below of=Step3X] (dot7) {};
    \node [dotlong, right of=dot7] (dot8) {};
    
    \path [linenoarrow] (EntityX) -- (dot1);
    \path [linenoarrow] (EntityY) -- (dot2);
     \path [linenoarrow] (dot1) -- (Step1X);
    \path [linenoarrow] (dot2) -- (Step2Y);
    \path [linenoarrow] (dot1) --node {$START$} (dot2);
    \path [line] (Step1X) --node {$\mathbf{u}$} (Step2Y);
    \path [line] (Step2Y) --node {$c_1, \mathbf{c}_2$} (Step3X);
    \path [linenoarrow] (Step2Y) -- (dot8);
    \path [linenoarrow] (Step1X) -- (Step3X);
    \path [linenoarrow] (Step3X) -- (dot7);
    \path [linenoarrow] (dot7) --node {$END$} (dot8);
\end{tikzpicture}
\caption{Flow diagram for the proposed lattice-based privacy-preserving scalar product computation for binary vectors} \label{Fig:PPSP Binary}
\end{figure}
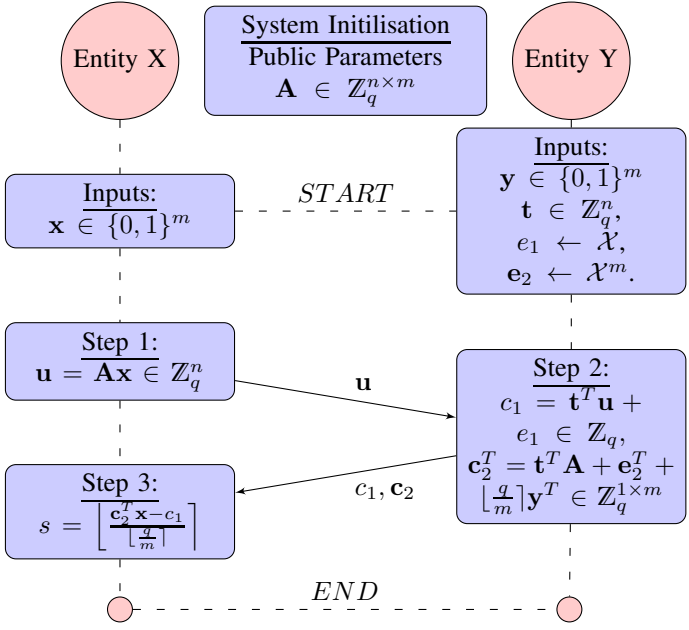

Let us suppose, there are two distrusting entities, X and Y. Entity X owns an $m-$dimensional binary vector $\mathbf{x} \in \{0,1\}^{m}$. Entity Y owns another $m-$dimensional binary vector $\mathbf{y} \in \{0,1\}^{m}$. Both X and Y want to interact with each other to compute the SP $s = \mathbf{x}^T\mathbf{y}$ without revealing their own vector to the other party. In the end, one-party obtains  $s = \mathbf{x}^T\mathbf{y}$. To perform PPSP using lattice, there are four steps required. The following subsections describe each of them in details.  The complete algorithm is given in Fig. \ref{Fig:PPSP Binary}.

\subsubsection{System initialisation} Let us start with generating  a uniformly random  matrix $\mathbf{A}  \in \mathds{Z}_q^{n \times m}$ which is known to X and Y. The matrix $\mathbf{A}$ contains  column vectors $\mathbf{a}_1$, $\mathbf{a}_2$, $\ldots$, $\mathbf{a}_m$ $\in \mathds{Z}_q^{n}$ i.e., $\mathbf{A}=[\mathbf{a}_1, \mathbf{a}_2, \ldots, \mathbf{a}_m ]$. 

\subsubsection{Step 1}
Entity X computes a SIS style vector using $\mathbf{A}$ and the binary vector $\mathbf{x}$ as 
\begin{equation}\label{Eq: Step 1 X computes Z1}
\mathbf{u} = \mathbf{Ax} ~(mod~q) \in \mathds{Z}_q^{n},
\end{equation}
and sends $\mathbf{u}$ to Y.

\subsubsection{Step 2}
Entity Y generates a uniformly random vector $\mathbf{t} \in \mathds{Z}_q^{n}$, a small error term $e_1  \leftarrow \mathcal{X}$, and a small error vector $\mathbf{e}_2 = [e_{2,1}, e_{2,2}, \ldots, e_{2,m}]^T \leftarrow \mathcal{X}^m$. Then Y computes the following LWE style term $c_1$ and vector $\mathbf{c}_2$:
\begin{eqnarray}
\label{Eq: Step 2 Y computes c1} c_1 &=& \mathbf{t}^T\mathbf{u}+ e_1 ~(mod~q) \in \mathds{Z}_q,\\
\label{Eq: Step 2 Y computes c2} \mathbf{c}_2^T &=& \mathbf{t}^T\mathbf{A}+ \mathbf{e}_2^T + \lfloor \frac{q}{m}\rceil\mathbf{y}^T ~(mod~q) \in \mathds{Z}_q^{1 \times m},
\end{eqnarray}
and sends these to X.

\subsubsection{Step 3}
Entity X performs the following computation to retrieve the SP value $s= \mathbf{x}^T\mathbf{y}$ as follows:
\begin{equation}\label{Eq: Step 3 X computes s}
s=\left\lfloor\frac{\mathbf{c}_2^T\mathbf{x} - c_1}{\lfloor \frac{q}{m}\rceil}\right\rceil.
\end{equation}

\subsection{Condition for Correctness}
Let us derive the condition for the above-mentioned algorithm to  output a correct result. In (\ref{Eq: Step 3 X computes s}),
\begin{eqnarray}
\nonumber \mathbf{c}_2^T\mathbf{x} - c_1 &=& (\mathbf{t}^T\mathbf{A}+ \mathbf{e}_2^T + \lfloor \frac{q}{m}\rceil\mathbf{y}^T)\mathbf{x} - (\mathbf{t}^T\mathbf{u}+ e_1),\\
\nonumber		  &=& \mathbf{t}^T\mathbf{A}\mathbf{x}+ \mathbf{e}_2^T\mathbf{x} + \lfloor \frac{q}{m}\rceil\mathbf{y}^T\mathbf{x} - \mathbf{t}^T\mathbf{u}- e_1.
\end{eqnarray}
Since $\mathbf{A}\mathbf{x} =\mathbf{u}$,  and $  \mathbf{t}^T\mathbf{A}\mathbf{x} = \mathbf{t}^T\mathbf{u}$,
\begin{eqnarray}
\label{eqn: SP with error} \mathbf{c}_2^T\mathbf{x} - c_1 &=& \lfloor \frac{q}{m}\rceil\mathbf{y}^T\mathbf{x} +\mathbf{e}_2^T\mathbf{x}  - e_1.
\end{eqnarray}
In (\ref{eqn: SP with error}), the scalar product is masked by error term $\mathbf{e}_2^T\mathbf{x}  - e_1$. To output a  correct answer, this error term must satisfy the following condition:
\begin{equation}\label{eqn: error condition}
    \mathbf{e}_2^T\mathbf{x}  - e_1 < \lfloor \frac{q}{2m}\rceil,
\end{equation}
hence,
\begin{equation}\label{eqn: error condition two}
    \frac{\mathbf{e}_2^T\mathbf{x}  - e_1}{\lfloor \frac{q}{m}\rceil} < \frac{1}{2}.
\end{equation}
 Therefore,
\begin{equation*}
\nonumber s = \left\lfloor \frac{\mathbf{c}_2^T\mathbf{x} - c_1}{\lfloor \frac{q}{m}\rceil} \right\rceil = \left\lfloor \frac{\lfloor\frac{q}{m}\rceil\mathbf{y}^T\mathbf{x} +\mathbf{e}_2^T\mathbf{x}  - e_1}{\lfloor \frac{q}{m}\rceil} \right\rceil =  \mathbf{y}^T\mathbf{x},
\end{equation*}
which proves the correctness of the proposed algorithm. Further, the requirements for the error term (\ref{eqn: error condition}) should be analysed and defined such that  $\mathbf{e}_2^T\mathbf{x}  - e_1$ is always smaller than $ \lfloor \frac{q}{2m}\rceil$. To achieve this, we need to find the upper bound for the error term. The following subsection is dedicated for this analysis.

\subsection{Upper bound of the error term ($ \mathbf{e}_2^T\mathbf{x}  - e_1 $)}\label{SS: UB for Binary}
{{ As we described in Section}} \ref{Subsection: LWE in intro},{{  the small error terms are sampled from a normal distribution with mean $0$ and standard deviation $\frac{\alpha}{\sqrt{2\pi}}$ (let us denote this as $\Psi_{0, \frac{\alpha}{\sqrt{2\pi}}}$) followed by scaling and modulo reduction by $q$  as follows:}}
\begin{equation}\label{eqn: sample e}
    e = \lfloor wq \rceil (mod~q)
\end{equation}
{{ where $w \leftarrow \Psi_{0, \frac{\alpha}{\sqrt{2\pi}}}$ and $e$ belongs to a ``rounded'' normal distribution with mean $0$ and standard deviation $\frac{\alpha q}{\sqrt{2\pi}}$ (let us denote this as $\mathcal{X}_{0,\frac{\alpha q}{\sqrt{2\pi}}}$). }}

Let us also denote  vectors $\mathbf{w} = [w_1, w_2, \ldots, w_m] \leftarrow \Psi^m_{0, \frac{\alpha}{\sqrt{2\pi}}}$ and $\bar{\mathbf{w}} = [w_1, w_2, \ldots, w_{m+1}] \leftarrow \Psi^{m+1}_{0, \frac{\alpha}{\sqrt{2\pi}}}$. Hence the error vector
\begin{equation}\label{eqn: sample error vector}
    \mathbf{e} = \lfloor \mathbf{w}q \rceil (mod~q).
\end{equation}
Using the above information, let us find the upper bound for the error term $\mathbf{e}_2^T\mathbf{x}  - e_1$. Let us define an $m+1$ dimensional vector $\bar{\mathbf{e}} = [\mathbf{e}_2^T~e_1]^T$ and another $m+1$ dimensional vector $\bar{\mathbf{x}} = [\mathbf{x}^T~-1]^T$, hence,  $\mathbf{e}_2^T\mathbf{x}  - e_1 = \bar{\mathbf{e}}^T\bar{\mathbf{x}}$. Using the triangle inequality, we can define the upper bound of the error term as follows:
\begin{equation}\label{eqn: norm UB}
|\mathbf{e}_2^T\mathbf{x}  - e_1| = |\bar{\mathbf{e}}^T\bar{\mathbf{x}}| \leq |(\bar{\mathbf{e}} -q\bar{\mathbf{w}})^T\bar{\mathbf{x}}| +  |(q\bar{\mathbf{w}})^T\bar{\mathbf{x}}|.
\end{equation}
Using the Cauchy-Schwarz inequality \cite{MatrixBook}, we can define the upper bound for the terms in (\ref{eqn: norm UB}) as follows:
\begin{eqnarray}
 |(\bar{\mathbf{e}} -q\bar{\mathbf{w}})^T\bar{\mathbf{x}}| &<&||\bar{\mathbf{e}} -q\bar{\mathbf{w}}||.||\bar{\mathbf{x}}||\\
 |(q\bar{\mathbf{w}})^T\bar{\mathbf{x}}|&<& ||q\bar{\mathbf{w}}||.||\bar{\mathbf{x}}||
\end{eqnarray}
According to (\ref{eqn: sample e}) and (\ref{eqn: sample error vector}), the rounding error for the components $w$ is at most $\frac{1}{2}$ (i.e., $e -\lfloor wq \rceil \leq \frac{1}{2}$), we have $||\bar{\mathbf{e}} -q\bar{\mathbf{w}}|| \leq \frac{\sqrt{m+1}}{2}$ and $||\mathbf{e}_1 -q\mathbf{w}|| \leq \frac{\sqrt{m}}{2}$.
 Hence,
\begin{equation*}
||\bar{\mathbf{e}} -q\bar{\mathbf{w}}||.||\bar{\mathbf{x}}|| + ||q\bar{\mathbf{w}}||.||\bar{\mathbf{x}}|| \leq \frac{\sqrt{m+1}}{2}||\bar{\mathbf{x}}|| + ||q\overline{\mathbf{w}}||.||\bar{\mathbf{x}}||.
\end{equation*}
Since $\bar{\mathbf{x}} \in \{0,1\}^{m+1}$, the Euclidean norm of   $\bar{\mathbf{x}}$ is $||\bar{\mathbf{x}}|| \leq \sqrt{m+1}$. Hence, 
\begin{equation*}
 \frac{\sqrt{m+1}}{2}||\bar{\mathbf{x}}|| + ||q\overline{\mathbf{w}}||.||\bar{\mathbf{x}}|| \leq \frac{m+1}{2} + ||q\overline{\mathbf{w}}||.\sqrt{m+1}.
\end{equation*}
Since $\overline{\mathbf{w}} \leftarrow \Psi_{0, \frac{\alpha}{\sqrt{2\pi}}}^{m+1}$ and $q\overline{\mathbf{w}} \leftarrow \mathcal{X}_{0, \frac{q\alpha}{\sqrt{2\pi}}}^{m+1}$, if we choose  standard deviation as $4.5$, then the probability 
\begin{equation*}
Pr\left(|qw| > 4.5 \times \frac{q\alpha}{\sqrt{2\pi}}\right) < 2.5 \times 10^{-7}, 
\end{equation*}
{{ (i.e., one in four million).  The probability will decrease further if we choose a higher number of standard deviations for the upper bound. Without loss of generality, in the rest of the paper, we consider  standard deviation as $4.5$. Therefore, with very high probability, }}
\begin{equation}\label{Equation: Std 4.5}
||q\bar{\mathbf{w}}|| \leq 4.5q\alpha\sqrt{\frac{m+1}{2\pi}}.  
\end{equation}
Therefore, with very high probability, the error
\begin{eqnarray}
\nonumber |\mathbf{e}_2^T\mathbf{x}  - e_1| &\leq& \frac{m+1}{2} + ||q\overline{\mathbf{w}}||.\sqrt{m+1},\\
\nonumber  &\leq& \frac{m+1}{2} + 4.5q\alpha\sqrt{\frac{m+1}{2\pi}}.\sqrt{m+1}.
\end{eqnarray}
As long as this error is smaller than $\lfloor \frac{q}{2m}\rceil$, i.e., 
\begin{eqnarray}
\label{Equation: UB for alpha binary}  \frac{m+1}{2} + 4.5q\alpha \frac{(m+1)}{\sqrt{2\pi}} &\leq& \left\lfloor \frac{q}{2m}\right\rceil,
\end{eqnarray}
our proposed solution outputs a correct result. Hence, if the upper bound for $\alpha$ is
\begin{eqnarray}
\label{Equation: condition 2 for alpha binary} \alpha &\leq& \frac{\sqrt{2\pi}}{4.5q(m+1)}\left[\lfloor \frac{q}{2m}\rceil - \frac{m+1}{2} \right],
\end{eqnarray}
then with high probability (it may not provide correct result one in four million times), the proposed algorithm outputs a correct result. This concludes the proof for correctness. The requirements for the correctness are listed in Table \ref{Table: Binary Parameters}. 

{{Extending the inputs from \{0,1\} to integer inputs \{0,1,2, …. l\} will lead to a smaller bin size i.e., $q/(m*l^2)$. Using this smaller size, the equations (14) to (23) can be revised to obtain parameters for input \{0,1,2, …. l\}. }} The next section analyses the security of the proposed algorithm.

\section{Security Analysis}\label{S: Security Analysis}
As defined in Section \ref{S: Binary PPSP} (refer to Fig. \ref{Fig:PPSP Binary}), the objective is to protect the privacy of $\mathbf{x}$ from Y and $\mathbf{y}$ from X. Entities X and Y interact with each other to compute the SP. 

{{ Firstly, let us prove that Y cannot learn the secret vector $\mathbf{x}$ from the exchanged vector $\mathbf{u}$ in Step 1.}} Since $\mathbf{x} \in \{0,1\}^m$ (therefore $\mathbf{x}$ is a short vector), according to the hardness of ISIS problem defined in Section \ref{Section: Lattice Hard Problems}, it is intractable for Y to solve $\mathbf{u} = \mathbf{Ax} ~mod~q$ and obtain a short vector as a solution. 

Step 1 operation is similar to hashing. Since the dimension of typical vector $\mathbf{x}$ is $10000$, there are $2^{10000}$ possibilities. The only problem is (as same as in any hashing algorithm) the output of Step 1 is deterministic for same $\mathbf{x}$. 

Therefore brute force approach may not work for Y. Hence Y needs to use mathematical properties to solve the problem to uncover $\mathbf{x}$ from $\mathbf{u}$. In other words, if Y can recover $\mathbf{x}$ from $\mathbf{u}$ then Y can solve the lattice hardest problem. As defined in Section \ref{Section: Lattice Hard Problems}, Y cannot find a vector $\mathbf{x}$ shorter than $\beta$ i.e., $||\mathbf{x}|| < \beta$. Therefore, let us analyse the shortest possible vector which can be recovered by Y. 

Suppose if Y wants to find a short vector  $\mathbf{x}$ from $\mathbf{u} = \mathbf{Ax} ~mod~q$ then Y may exploit the state-of-the-art techniques called lattice reduction method \cite{gamaandnguyen}  and/or combinatorial method  \cite{CombinotorialBlum}. Denote the shortest vector which can be found by these techniques as $\mathbf{x}_s$. It is proven in literature (theoretically and experimentally)\cite{LatticeBasedcrypto}, that the Euclidean length of $\mathbf{x}_s$ has a lower-bound as follows:
\begin{equation}\label{Equation: shortest vector}
    ||\mathbf{x}_s|| \geq min\left\{q, 2^{2\sqrt{n.log(q)log(\delta)}}\right\},
\end{equation}
where $\delta \geq 1.01$ \cite{gamaandnguyen}. Since the X's secret vector $\mathbf{x} \in \{0,1\}^m$, the Euclidean length $||\mathbf{x}|| \leq \sqrt{m}$. Hence, using (\ref{Equation: shortest vector}) and assuming $q$ is very large, if 
\begin{equation}\label{Equation: upperbound for m}
\sqrt{m} <  2^{2\sqrt{n.log(q)log(\delta)}},
\end{equation}
 then Y cannot recover $\mathbf{x}$ from $\mathbf{u}$. 
This is a first condition for security. This concludes that if condition (\ref{Equation: upperbound for m}) is met then Y cannot recover $\mathbf{x}$ from $\mathbf{u}$. 
Also, the cost ($L$) of finding a short binary vector using the  techniques described above is defined as \cite{LatticeBasedcrypto}:
\begin{equation}\label{Equation: cost shortest vector}
    L \approx 2^{\frac{m}{2^k}},
\end{equation}
where   $k$ should satisfy the following equation:
\begin{equation}\label{Equation: cost shortest vector k}
   \frac{2^k}{k+1} \approx \frac{m}{n.log(q)}.
\end{equation}
Now let us focus whether X can recover $\mathbf{y}$ from the messages $c_1$ and $\mathbf{c}_2$ sent by Y to X in Step 2.

\begin{table*}[!ht]\centering
\caption{Requirements for Parameters to Achieve 128-bit security and Correctness when the standard deviation is set for 4.5.}
\begin{tabular}{|c|c|c|}\hline
                & $n$     & $m$              \\ \hline
Correctness &  $n \geq 1$       &   $m \geq 1$       \\ \hline
Security    & $n.log(q)>128$ & $m \geq nlog(q)$  \& $\sqrt{m} <  2^{2\sqrt{nlog(q)log(\delta)}}$     \\ \hline \hline
                          & $\alpha$   & $q$    \\ \hline 
Correctness &        $\alpha \leq \frac{\sqrt{2\pi}}{4.5q(m+1)}\left[\lfloor \frac{q}{2m}\rceil - \frac{m+1}{2} \right]$ & $q > 2m$ \\ \hline
Security    &  $\alpha \geq max\left\{\frac{2\sqrt{n}}{q}, 1.5\sqrt{2\pi}.max\left\{ 1/q, 2^{-2\sqrt{n.log(q).log(\delta)}} \right\}\right\}$                                                                               & $q > n$     \\ \hline
\end{tabular}\label{Table: Binary Parameters}
\end{table*}


According to the definition in Section \ref{Section: Lattice Hard Problems}, if $c_1$ and $\mathbf{c}_2$ are LWE terms then it is intractable  for X to recover $\mathbf{y}$ since $c_1$ and $\mathbf{c}_2$ are indistinguishable from uniformly random distribution. If $\mathbf{t}$, $\mathbf{u}$, and $\mathbf{A}$ are uniformly distributed and the error term $e_1$ and error vector $\mathbf{e}_2$ are sampled from normal distribution with standard deviation greater than $2\sqrt{n}$ as defined in (\ref{Equation: condition 1 for alpha}) then $c_1$ and $\mathbf{c}_2$ are uniformly random. 

Matrix $\mathbf{A}$ is already a uniformly random matrix. Entity Y can generate uniformly random $\mathbf{t}$, $e_1$ and  $\mathbf{e}_2$. The vector $\mathbf{u}$ sent by X is uniformly random  as long as the number of possibilities for $\mathbf{x}$ is larger than $\mathbf{u}$ i.e., $2^m > q^n$ or $m > n.log(q)$ \cite{LatticeBasedcrypto} (this is the second security condition). 

Since the dimension of $\mathbf{t}$ is $m > 1$, and the  scalar $\mathbf{t}^T\mathbf{u}$ is masked by an error term $e_1$, the term $c_1$ is scalar and completely random. Therefore, according to the LWE definition, it is intractable for X to recover the elements of $\mathbf{t}$ from  scalar $c_1$. To analyse $\mathbf{c}_2$, let us denote the $i$th element of $\mathbf{c}_2$ as $c_{2,i}$ where $c_{2,i} = \mathbf{t}^T\mathbf{a}_i + e_{2,1} + \lfloor \frac{q}{2m}\rceil y_i$. In $c_{2,i}$, $\mathbf{t}^T\mathbf{a}_i + e_{2,1}$ is scalar and LWE term i.e., uniformly random. Similar to LWE encryption scheme \cite{RegevLWE2005}, $\mathbf{t}^T\mathbf{a}_i + e_{2,1}$  acts like a one-time pad to hide the message $ \lfloor \frac{q}{2m}\rceil y_i$. Hence, X cannot recover $y_i$ from $c_{2,i}$ and therefore the proposed scheme is secure. In Section \ref{Section: Parameter Selection}, we show that our parameter choice satisfying (\ref{Equation: condition 1 for alpha}) (third security condition) is hard and at least equivalent to $128-$bit security.

In LWE, the noise term plays a major role in determining the hardness \cite{RegevLWE2005}. The normal distribution where the error terms are sampled must  satisfy (\ref{Equation: condition 1 for alpha}). The $\alpha$ term must be chosen as largest possible while satisfying (\ref{Equation: condition 1 for alpha}) for hardness of LWE. To quantify the hardness or security level of LWE for a concrete set of parameters, Regev et. al  exploited the dual lattice in \cite[p.~21]{LatticeBasedcrypto}. The idea is to find how many operations are required to distinguish an LWE term from uniform distribution. This is only possible if an adversary can find a short vector on dual lattice. To this, let us denote  a vector  $\mathbf{v}$ and denote a short vector in dual lattice as $\mathbf{w}$. If  the vector  $\mathbf{v}$  is an LWE vector then the scalar product   $\mathbf{v}^T\mathbf{w}$ will be an integer \cite[p.~22]{LatticeBasedcrypto}. If not then   $\mathbf{v}$ is a uniform random vector.  Therefore finding a short vector in dual lattice must be hard. If the standard deviation of the error term $\alpha q/2\pi$ is not bigger than $1/||\mathbf{w}||$ then it may be possible to find a short vector in dual lattice. Therefore, error term must be bigger than $1/||\mathbf{w}||$ for LWE security. This requirement and (\ref{Equation: shortest vector}) can now be used to quantify the LWE security. 

Now using the lattice properties i.e., the length of a shorter vector in dual lattice is equivalent to $1/q$ times the  length of shorter vector in  lattice  \cite[p.~22]{LatticeBasedcrypto}. Using this  and (\ref{Equation: shortest vector}), we can say $||\mathbf{w}|| \approx \frac{1}{q}.min\left\{q, 2^{2\sqrt{n.log(q)log(\delta)}}\right\}$. Therefore if error 
\begin{equation}\label{Equation: LWE alpha term}
\frac{\alpha q}{\sqrt{2\pi}} >> \frac{1}{||\mathbf{w}||},
\end{equation}
then LWE is hard. By taking $1.5$ as factor, we can define the lower-bound for $\alpha$ from (\ref{Equation: LWE alpha term}) as follows \cite{LatticeBasedcrypto}:
\begin{equation}\label{Equation: LWE lowerbound for alpha term}
\alpha \geq 1.5\sqrt{2\pi}.max\left\{ 1/q, 2^{-2\sqrt{n.log(q).log(\delta)}} \right\}.
\end{equation}
The cost of finding a shorter vector is same as (\ref{Equation: cost shortest vector}). In Section \ref{Section: Parameter Selection}, we show that our parameter choice to satisfy (\ref{Equation: condition 1 for alpha}) is hard and at least equivalent to $128-$bit security.



\subsection{Parameter Selection}\label{Section: Parameter Selection}
Firstly, let us obtain the relationship between $q$ and  $m$. Since the maximum possible value for $\mathbf{x}^T\mathbf{y}$ is $m$, we split $q$ into $m$ parts i.e., the distance between the consecutive  values is $\lfloor \frac{q}{m} \rceil$. To obtain a correct result, as shown in (\ref{Equation: UB for alpha binary}), half of this distance should be larger  to accommodate the error term i.e., $\lfloor \frac{q}{2m} \rceil > 1$ or $q > 2m$.  Table \ref{Table: Binary Parameters}  provides the necessary requirements for  all the parameters to achieve correctness and security. This table is a summary of requirements derived in the previous sections. Using this table, let us obtain a concrete set of parameters  to achieve $128-$bit security. The same strategy has been used to obtain the parameters for lower security (i.e., $80-$bits, and $112-$bits) and higher security $256-$bits in Section \ref{Section: Experiments}.

To obtain $128-$bit security, we need to choose our parameters in such a way that the cost equation (\ref{Equation: cost shortest vector}), $L \approx 2^{\frac{m}{2^k}} \geq 2^{128}$.  If we choose $k=2$ then from (\ref{Equation: cost shortest vector k}), $m \approx n.log(q)$. Hence, $L \approx 2^{n.log(q)} \geq 2^{128}$. Therefore the security of the solution would be equal to $128-$bits if $n.log(q) \approx m \geq 128$. Based on this and other requirements (all are listed in Table \ref{Table: Binary Parameters}), we are proposing six sets of parameters in Table \ref{Tabe: Parameters} to achieve $128-$bits security. These parameters have been cross validated using the well known LWE Estimator \cite{LWEEstimator} [- the source code for the LWE Estimator, that calculates the security complexity using six different algorithms such as lattice-reduction, dual-lattice attacks etc, is available at https://bitbucket.org/malb/lwe-estimator].

\begin{table}[!ht]\centering
\caption{Choices for the security parameters to achieve at least $128-$bit Security.}
\begin{tabular}{|c|c|c|c|c|c|}
\hline
SET    & $n$    & \begin{tabular}[c]{@{}c@{}}$m$\\ $\approx$\end{tabular} & \begin{tabular}[c]{@{}c@{}}$q$\\ $\approx$\end{tabular} & \begin{tabular}[c]{@{}c@{}}$Security$\\ $\approx$\end{tabular} & \begin{tabular}[c]{@{}c@{}}$\alpha .q$\\ (error std. $\approx$)\end{tabular} \\ \hline
I   & $50$   & $2^{15}$                                                & $2^{570}$                                               & $2^{128}$                                                      & $2^{538}$                                                                    \\ \hline
II  & $100$  & $2^{15}$                                                & $2^{270}$                                               & $2^{128}$                                                      & $2^{238}$                                                                    \\ \hline
III & $250$  & $2^{15}$                                                & $2^{116}$                                               & $2^{128}$                                                      & $2^{85}$                                                                     \\ \hline
IV  & $500$  & $2^{15}$                                                & $2^{55}$                                                & $2^{128}$                                                      & $2^{24}$                                                                     \\ \hline
V   & $1000$ & $2^{15}$                                                & $2^{39}$                                                & $2^{187}$                                                      & $2^7$                                                                        \\ \hline
VI  & $2000$ & $2^{16}$                                                & $2^{41}$                                                & $2^{517}$                                                      & $2^7$                                                                        \\ \hline
\end{tabular}\label{Tabe: Parameters}
\end{table}

In Table \ref{Tabe: Parameters}, parameters $n$ and $q$ play a major role to ensure $128-$bit security. They are linked as increasing  $n$ leading to a small $q$. These  parameters determine the size of matrix $\mathbf{A}$ and  the memory requirement. The first four sets are equivalent in terms of memory ($\approx 100MB$) while the last two require around $200MB$ and $800MB$, respectively. As shown in the experiments, running time for the last two   are significantly higher and not useful for practical applications. For Sets V and VI, the size of $q$ is not decreasing as much as those for the other sets. The security levels for Sets V and VI are $187-$bits and $517-$bits, respectively. {{  The reason is that, larger $n$ leads to a larger $m$, hence, in order to satisfy the error distribution parameter $\alpha$ in}} (\ref{Equation: condition 2 for alpha binary}), {{the value for $q$ must be set to high. Increasing the value for $\alpha$ will increase the security.}}


\section{Experimental Results}\label{Section: Experiments}
In order to evaluate the proposed LWE based PPSP scheme, we implemented the algorithm in Java and tested on a 64-bit Windows PC with 16GB RAM and Intel(R) Core(TM) i5-4210U CPU at 1.70GHz. For performance comparison, we also implemented the Paillier homomorphic encryption based PPSP scheme \cite{Paillier} on the same PC using Java. Additionally, we compared our scheme with one of the most efficient PPSP algorithms in \cite{SPOC2}. Our test results show that the proposed LWE based scheme is significantly faster (at least $10^5$ times faster) than the Paillier homomorphic PPSP scheme and at least twice as fast as  \cite{SPOC2} for the $128-$bit security. 

\begin{table}[!ht]\small
\caption{Paillier homomorphic encryption based PPSP \cite{Paillier}.}
  \centering
  \begin{tabular}{|l|}\hline
\textbf{Input~by X:}   $\mathbf{a}=[a_{1},\ldots,a_{m}]^T$ $\in$ $\{0,1\}^m$\\
~~~ and \textbf{Y:} $\mathbf{b}=[b_{1},\ldots,b_{m}]^T$ $\in$ $\{0,1\}^m$\\
\textbf{Output to X:} $\mathbf{a}^T\mathbf{b}$ \\ \hline
\textbf{Step 1: X performs the following operations:}\\
~~Generates Paillier public-private key pairs $\{pub,sk\}$,\\
~~~ ~~FOR EACH $a_i$, $i=1,2,..., m$\\
~~~ ~~~~~~Computes\\
~~~ ~~~~~~~~~ $E_{pub}(a_i) = \llbracket a_i \rrbracket$,\\
~~~ ~~END FOR\\
~~keeps $sk$, and sends $(pub, E_{pub}(a_1) \ldots E_{pub}(a_m))$ to Y\\ \hline
\textbf{Step 2: Y executes the following operations}\\
~~~ ~~Using $b_i$, $i=1,2,..., m+2$\\
~~~ ~~~~~~Computes\\
~~~ ~~~~~~~~~ $E(\mathbf{a}^T\mathbf{b}) = \llbracket a_1 \rrbracket^{b_1}.\llbracket a_2 \rrbracket^{b_2}\ldots \llbracket a_m \rrbracket^{b_m}$\\
~~~ Sends $E(\mathbf{a}^T\mathbf{b})$ back to X\\ \hline
\textbf{Step 3: X decrypts  and obtains}\\
~~~~~~~ $\mathbf{a}^T\mathbf{b}$ =  $ D_{sk}(E(\mathbf{a}^T\mathbf{b}))$. \\ \hline
\end{tabular}
\label{Table: Paillier}
\end{table}

\subsection{Proposed Lattice-based PPSP Scheme and Paillier PPSP scheme}
The Paillier cryptosystem \cite{Paillier} is an additively homomorphic public-key encryption scheme. Its  provable
semantic security is based on the decisional composite residuosity problem:  it is mathematically intractable to decide whether an integer $z$ is an $n$-residue modulo $n^2$ for some composite $n$, i.e. whether there exists some $y \in \mathcal{Z}^*_{n^2}$ such that $z=y^n \mod n^2$.  Let $n=pq$ where $p$ and $q$ are two large prime numbers. A message $m \in \mathcal{Z}_n$ can be encrypted using the Paillier cryptosystem as $\llbracket m \rrbracket=g^mr^n~\textrm{mod}~n^2$ where $g\in \mathcal{Z}^*_{n^2}$ and $r\in \mathcal{Z}^*_{n}$. For a given encryption $\llbracket m_1 \rrbracket$ and $\llbracket m_2 \rrbracket$, an encryption $\llbracket m_1+m_2\rrbracket$ can be obtained as $\llbracket m_1+m_2\rrbracket=\llbracket m_1\rrbracket\llbracket m_2\rrbracket$, and multiplication of an encryption $\llbracket m_1\rrbracket$ with a constant $\alpha$ can be computed efficiently as $\llbracket m_1.\alpha\rrbracket=\llbracket m_1\rrbracket^\alpha$. Hence,  a Paillier cryptosystem is an additively homomorphic cryptosystem. Let us denote  $E()$ and $D()$ as the Paillier homomorphic encryption and decryption functions.  Using the homomorphic properties and the above definitions, homomorphic encryption based PPSP is described in Table \ref{Table: Paillier}.

According to NIST recommendation \cite{NIST1,NIST2}, public-key encryption schemes such as RSA and Paillier must use $3072-$bit long keys for encryption and decryption in order to achieve $128-$bit security. Hence, to obtain the running time for the Paillier homomorphic encryption based PPSP, we used $3072-$bit long  keys. We also obtained the running time for the proposed LWE based scheme for the first five sets of parameter given in Table \ref{Tabe: Parameters} (Sixth set was ignored as it was taking too much time to run). The running times averaged over 100 executions are listed  in Table \ref{Table: Time comparision for 128 bit security} [no parallelization or multi-threading was used].

\begin{table}[!ht]\centering\small
\centering
\caption{Average running time for the proposed and Paillier-based PPSP schemes.}
\begin{tabular}{|c|c|c|c|c|c|}
\hline
\multirow{2}{*}{} & \multicolumn{4}{c|}{\begin{tabular}[c]{@{}c@{}}The Proposed \\ Lattice-based PPSP\end{tabular}}                                                                                                                                       & \multirow{2}{*}{\begin{tabular}[c]{@{}c@{}}Pailler Based\\ PPSP \\ (ms)\end{tabular}} \\ \cline{2-5}
          SET        & \begin{tabular}[c]{@{}c@{}}Step 1\\ (ms)\end{tabular} & \begin{tabular}[c]{@{}c@{}}Step 2\\ (ms)\end{tabular} & \begin{tabular}[c]{@{}c@{}}Step 3\\ (ms)\end{tabular} & \textbf{\begin{tabular}[c]{@{}c@{}}Total\\ (ms)\end{tabular}} &                                                                                       \\ \hline
I                 & $692$                                                 & $2482$                                                & $21$                                                  & \textbf{$3195$}                                               & $\approx 5\times 10^8$                                                                \\ \hline
II                & $756$                                                 & $3207$                                                & $9$                                                   & \textbf{$3972$}                                               & $\approx 5\times 10^8$                                                                \\ \hline
III               & $2456$                                                & $7146$                                                & $12$                                                  & \textbf{$9614$}                                               & $\approx 5\times 10^8$                                                                \\ \hline
IV                & $4721$                                                & $16972$                                               & $9$                                                   & \textbf{$21702$}                                              & $\approx 5\times 10^8$                                                                \\ \hline
V                 & $129328$                                              & $206741$                                              & $8$                                                   & \textbf{$336077$}                                             & $\approx 8\times 10^8$                                                                \\ \hline
\end{tabular}\label{Table: Time comparision for 128 bit security}
\end{table}
As presented in Table \ref{Table: Time comparision for 128 bit security}, the result of Set I has outperformed the other sets. This is due to the fact that, even though the security levels are equal across all the sets, when the size for $n$ increases, the matrix $\mathbf{A}$ becomes larger and requires an increased number of multiplications. In turn, this slows down the algorithm. With this observation, we will continue using the parameters that belong to Set I for the remainder of our experiments presented in this paper. The last column in Table \ref{Table: Time comparision for 128 bit security} shows the average running time for the Paillier scheme. The proposed scheme is at least $10^5$ times faster than Paillier PPSP scheme. The dimensions of the input vectors for these sets are in the range of $20000$ to $50000$ (see the third column in Table \ref{Tabe: Parameters}).

\begin{figure}\centering
\includegraphics[trim={0cm 0cm 0cm 0cm},clip, width=3.4in]{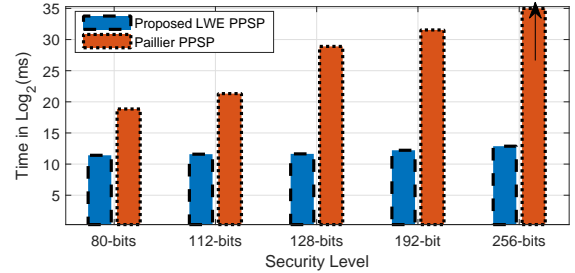}
\caption{{{ Average running time for the proposed LWE PPSP scheme against the Paillier PPSP scheme for different security levels. Note that y-axis is in log scale.}}}
\label{Fig: LWEvsPaillier}
\end{figure}

To compare the performance of the proposed scheme for different security levels, a new set of parameters are provided in Table \ref{Table: Parameters for LWE Paillier}. Based on the NIST recommendations \cite{NIST1,NIST2},  the key sizes for the Paillier scheme is also provided in Table \ref{Table: Parameters for LWE Paillier}. Using this information, the average running time is plotted in Fig \ref{Fig: LWEvsPaillier}. While the average running time for the proposed scheme is increasing linearly, it increases exponentially for the Paillier scheme. It should be noted that the average running time for the proposed scheme is around $8$ seconds at $256-$bit security [without any parallel computations or multi-threading]. These results demonstrate that the proposed lattice PPSP scheme is significantly faster than the Paillier PPSP.

\begin{table}[!ht]\centering
\centering
\caption{Parameters and key sizes for the proposed and Paillier based PPSP schemes for different levels of security.}
\begin{tabular}{|c|c|c|c|c|c|}
\hline
Security  & $n$  & \begin{tabular}[c]{@{}c@{}}$m$\\ $\approx$\end{tabular} & \begin{tabular}[c]{@{}c@{}}$q$\\ $\approx$\end{tabular} & \begin{tabular}[c]{@{}c@{}}$\alpha .q$\\ $\approx$\end{tabular} & \begin{tabular}[c]{@{}c@{}}Paillier\\ Key Size\end{tabular} \\ \hline
$2^{80}$  & $50$ & $23500$                                                 & $2^{470}$                                               & $2^{439}$                                                       & $1024$                                                      \\ \hline
$2^{112}$ & $50$ & $27500$                                                 & $2^{550}$                                               & $2^{518}$                                                       & $2048$                                                      \\ \hline
$2^{128}$ & $50$ & $28500$                                                 & $2^{570}$                                               & $2^{538}$                                                       & $3072$                                                      \\ \hline
$2^{192}$ & $50$ & $40500$                                                 & $2^{810}$                                               & $2^{777}$                                                       & $7680$                                                      \\ \hline
$2^{256}$ & $50$ & $50000$                                                 & $2^{1000}$                                              & $2^{997}$                                                       & $15360$                                                     \\ \hline
\end{tabular}\label{Table: Parameters for LWE Paillier}
\end{table}

\subsection{Proposed Scheme and Randomisation Technique}
 Table \ref{Table: SPOC} shows the state-of-the-art randomisation based  PPSP  \cite{Rahul_TASL,SPOC2}. The security of this algorithm depends on the hardness of the factoring an integer i.e., $C_i=s(a_i.\alpha + c_i) ~\mathrm{mod}~ p,~ a_i \neq 0$. $C_i$s are protected by $s$ and known only to X. If Y wants to recover the X's input vector, Y needs to factor all $C_i$s to find the common $s$. This approach can be seen as an approach used in RSA encryption or any public-key encryption that relies on hardness of factoring integers. According to the NIST recommendation \cite{NIST1,NIST2}, the size of these integers must be around $3072-$bit in order to obtain $128-$bit security (without loss of generality, we ignore the requirement of prime numbers). Hence, we set $k1$ in Table \ref{Table: SPOC} to $3072-$bits to compare randomisation-based PPSP and the proposed lattice PPSP scheme. 

\begin{figure}\centering
\includegraphics[trim={0cm 0cm 0cm 0cm},clip, width=3.4in]{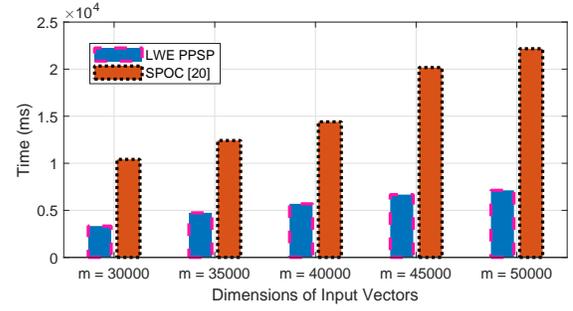}
\caption{{{ Average running time for the proposed LWE PPSP scheme against the Randomisation-based PPSP scheme}} \cite{Rahul_TASL,SPOC2} for different sizes of input vectors.}
\label{Fig: LWEvsSPOC}
\end{figure}

Using this setting, the average running time for the proposed and randomisation based PPSP schemes are obtained at $128-$bit security. Fig. \ref{Fig: LWEvsSPOC} shows the average running times for both  schemes for different input vectors whose dimensions are between $30000$ and $50000$. The proposed scheme is at least twice as fast compared to randomisation based scheme for the security parameters. It should be noted that, since randomisation-based scheme relies on hardness of integer factorisation,  similar to Paillier scheme, it is also vulnerable for quantum attacks.

\begin{table}[!ht]\small
\caption{Randomisation based PP scalar product algorithm.}
  \centering
  \begin{tabular}{|l|}\hline
\textbf{Input~by X:}   $\mathbf{a}=[a_{1},\ldots,a_{m}]^T$ $\in$ $\{0,1\}^m$\\
~~~ and \textbf{Y:} $\mathbf{b}=[b_{1},\ldots,b_{m}]^T$ $\in$ $\{0,1\}^m$\\
\textbf{Output to X:} $\mathbf{a}^T\mathbf{b}$ \\ \hline
\textbf{Step 1: X performs the following operations:}\\
~~Given security parameters $k_1$, $k_2$, $k_3$, $k_4$,\\
~~~~~~choose two large primes $\alpha$, $p$\\
~~~~~~such that $|p| = k_1$, $|\alpha|=k_2$, set $a_{m+1}=a_{m+2}=0$\\
~~Choose a large random number $s \in Z_p$, and $m+2$ random\\
~~~ ~~ numbers $c_i$, $i=1,2,..., m+2$, with $|c_i|=k_3$\\
~~~ ~~FOR EACH $a_i$, $i=1,2,..., m+2$\\
~~~ ~~~~~~Compute\\
~~~ ~~~~~~~~~ $C_i=s(a_i.\alpha + c_i) ~\mathrm{mod}~ p,~ a_i \neq 0$\\
~~~ ~~~~~~~~~ $C_i=sc_i ~\mathrm{mod}~ p,~ a_i = 0$\\
~~~ ~~END FOR\\
~~keeps $s^{-1} \mathrm{mod} ~p$ secret, and sends $(\alpha, p, C_1 \ldots C_{m+2})$ to Y\\ \hline
\textbf{Step 2: Y executes the following operations}\\
~~~ ~~set $b_{m+1}=b_{m+2}=0$\\
~~~ ~~FOR EACH $b_i$, $i=1,2,..., m+2$\\
~~~ ~~~~~~Compute\\
~~~ ~~~~~~~~~ $D_i=b_i.\alpha.C_i~\mathrm{mod}~p,~ b_i \neq 0$\\
~~~ ~~~~~~~~~ $D_i=r_i.C_i ~\mathrm{mod}~ p,~ b_i = 0$,\\
~~~ ~~~~~~~~~ where $r_i$ is a random number with $|r_i| = k_4$\\
~~~ ~~END FOR\\
~~~ Send $D = \sum_{i=1}^{m+2} D_i ~\mathrm{mod}~p$  to X\\ \hline
\textbf{Step 3: Now X computes  and obtains}\\
~~$E=s^{-1}.D~\mathrm{mod}~p$ and  get $\mathbf{a}^T\mathbf{b}$\\
~~ $ = \sum_{i=1}^{n}a_i.b_i = \frac{E-(E~\mathrm{mod}~\alpha^2)}{\alpha^2}$. \\ \hline
\end{tabular}
\label{Table: SPOC}
\end{table}

 Even though the proposed scheme is developed to protect the PP applications against the quantum computers, the efficiency analysis shows that the algorithm can be used to replace the existing schemes. Running time in Table \ref{Table: Time comparision for 128 bit security} is obtained from sequential programming. It is taking around $3$ seconds to execute the SP of two vectors whose dimensions are around $30000$. Nearly $2.5$ seconds are spent on Step 2 calculating (\ref{Eq: Step 2 Y computes c2}). This equation can be computed in parallel i.e.,  $\mathbf{t}^T\mathbf{A}$  is equivalent to $\mathbf{t}^T\mathbf{a}_i$ where $i \leq m$. Therefore, we used multi threading features of Java to speed-up the process. By setting four threads, average running time has been reduced to $1.2$ seconds from $3$ seconds.

\subsection{Communication Complexity}
{{ Using the algorithms in}} Fig. \ref{Fig:PPSP Binary} (the proposed LWE scheme), Table \ref{Table: Paillier} (Paillier Homomorphic Encryption Scheme based PPSP), and Table \ref{Table: SPOC} (Randomisation based PPSP), {{  we can calculate the communication cost in terms of transmitted bits between Entity \textbf{X} and Entity \textbf{Y}. }}
\subsubsection{Total bits transmitted from Entity X to Entity Y} 
{{ Total number of bits required to for the proposed LWE based PPSP scheme is $n*log_2(q)$. Similarly, $m*log_2(pub)$ and $(m+4)*log_2(k1)$ number of bits are required for the Paillier based scheme and Randomisation scheme, respectively.}}

\subsubsection{Total bits transmitted from Entity Y to Entity X} 
{{ Total number of bits required to for the proposed LWE based PPSP scheme is $(m+1)*log_2(q)$. Similarly, $log_2(pub)$ and $log_2(k1)$ number of bits are required for the Paillier based scheme and Randomisation scheme, respectively.}}

{{ At 128-bit level security, if we extract the parameters, then $n=50$, $log_2(q) = 570$, $log_2(pub)=3072$, and $log_2(k1)=3072$. Using these parameters, Table}} \ref{Table:Communication cost} {{ shows the communication cost for all three schemes when the dimension of the input vectors is $m=30000$. It’s clear from Table 7 that the LWE scheme significantly benefits from a shorter prime number (six times smaller than the other schemes’ prime number) and achieves six times lower data requirement to perform the scalar computation.}}

\begin{table}[!ht]\centering
\centering
\caption{{{ Communication cost comparison.}}}
\begin{tabular}{|c|c|c|c|}
\hline
                            & \textbf{X to Y} & \textbf{Y to X} & \textbf{Total} \\ \hline
\textbf{Proposed LWE PPSP}  & 3.6 kB               & 2.1 MB               & $\sim$2 MB     \\ \hline
\textbf{Paillier PPSP}      & 11.5 MB              & 0.3 kB               & $\sim$12 MB    \\ \hline
\textbf{Randomisation PPSP} & 11.5 MB              & 0.3 kB               & $\sim$12 MB    \\ \hline
\end{tabular}\label{Table:Communication cost}
\end{table}

\section{Conclusions and Future Work}\label{Section: Conclusions}
In this paper  a novel privacy-preserving scalar product computations using the fundamentals of lattice-based cryptography has been proposed. In particular, the proposed scheme was built directly on top of the lattice hard problems such as shortest integer solution and learning with errors.  $128-$bit encryption security has been achieved with the proposed framework. Several validation and verification experiments have shown that the proposed scheme is one of the best performing scheme in terms of complexity whilst not compromising systems security. 

\subsection*{Challenges and Future Work}
The dimensions of the input vectors  depend on $n$ and $q$ i.e., $m = n.log_2(q)$. Hence the proposed work supports larger dimensions such as $30000$. Even though, this is appropriate for many applications,  converting the solution to support smaller dimensions such as $100$ would be an interesting problem that requires further investigations. 

\balance

\end{document}